\documentclass[a4paper,11pt]{article}
\pdfoutput=1 

\usepackage{jcappub} 

\usepackage[T1]{fontenc} 
\usepackage{slashed}
\usepackage{multicol}
\usepackage[compat=1.1.0]{tikz-feynman}
\usepackage{subcaption}

\title{\boldmath Freeze-in leptogenesis with sterile neutrino self-interactions}

\author{Mar\'ia Dias Astros and}
\author{Stefan Vogl}

\affiliation{Institute of Physics, University of Freiburg,\\ Herrmann-Herder-Str. 3, 79104 Freiburg, Germany}

\emailAdd{maria.dias@physik.uni-freiburg.de}
\emailAdd{stefan.vogl@physik.uni-freiburg.de}

\abstract{Sterile neutrinos are a simple yet compelling addition to the Standard Model. 
For right-handed neutrinos with masses below the electroweak scale, leptogenesis can proceed through CP-violating oscillations of the sterile neutrinos. This is known as ARS or freeze-in leptogenesis. 
However, the ARS scenario requires the right-handed neutrinos to have a high degree of mass degeneracy. 
In this work, we study an extension of the SM that introduces a scalar singlet in addition to the two sterile neutrinos required to generate the baryon asymmetry. The new scalar interacts with the sterile neutrinos via a Yukawa interaction. 
This leads to an additional rate for the production and destruction of the sterile neutrinos and to a novel contribution to the effective potential. 
For the case in which the mass and the new Yukawa matrices are not diagonal in the same basis, we find that the effective potential can boost the baryon asymmetry of the universe by several orders of magnitude. This significantly alleviates the fine-tuned mass condition required in vanilla ARS leptogenesis.    }

\begin{document}
\maketitle
\flushbottom

\section{Introduction}
\label{sec:intro}

The Standard Model of particle physics (SM) is widely regarded as one of the most successful theories in modern physics. However, it is well-established that the SM fails in explaining various phenomena, including  dark matter, dark energy, neutrino masses, and the baryon asymmetry of the universe. Thus, the SM remains an incomplete theory of nature. Sterile neutrinos, as one of the most minimal and compelling extensions of the SM, are promising candidates for addressing several of its key shortcomings. They have the potential to provide insights into the origin of neutrino masses, the nature of dark matter, and the baryon asymmetry of the universe (BAU). The latter refers to the Standard Model's inability to account for the observed matter-antimatter imbalance that is quantified by the baryon-to-entropy ratio\footnote{It is also common to quantify the BAU by the baryon-to-photon ratio $\eta = (n_{b} - n_{\bar{b}})/  {n_{\gamma}} \approx 6 \times 10^{-10}$ with $n_{ \gamma}$ the number density of photons.} \cite{Planck:2018vyg,ParticleDataGroup:2024cfk,Yeh:2022heq} 
\begin{equation*}
    \frac{B}{s} = \frac{n_{b}-n_{\bar{b}}}{s} \approx 8 \times 10^{-11},
    \label{eq: value of B}
\end{equation*}
where $s$ is the entropy density of the universe and $n_{b,\bar{b}}$ are the number densities of baryons and antibaryons, respectively. Leaving aside the possibility that the BAU is the result of fine-tuned initial conditions, numerous ideas have been put forward to give a dynamical explanation for the observed value. Among these, leptogenesis \cite{Fukugita:1986hr}, has become a very active field of research, particularly in the wake of the discovery of neutrino oscillations. In these scenarios, the BAU is produced in the leptonic sector and subsequently transferred to baryons via sphalerons \cite{Kuzmin:1985mm}. Leptogenesis is specially compelling because it appears as a generic mechanism to generate the BAU in type I seesaw scenarios that also account for the lightness of the active neutrinos. In its original implementation, the SM is extended by the introduction of sterile neutrinos with masses well above the electroweak scale. These are responsible for generating the BAU through their out-of-equilibrium decays. This mechanism is known as thermal or freeze-out leptogenesis, as the right-handed neutrinos (RHNs) (produced during reheating) are initially in thermal equilibrium with the SM plasma. Nonetheless, models of leptogenesis with lighter sterile neutrinos, potentially accessible in current and future experiments, are very intriguing. These include, e.g. the neutrino minimal SM ($\nu$MSM) \cite{Asaka:2005pn, Asaka:2005an, Shaposhnikov:2008pf, Asaka:2010kk, Canetti:2010aw} and resonant leptogenesis \cite{Pilaftsis:1997jf, Pilaftsis:2003gt}. The $\nu$MSM, in particular, makes use of an idea proposed by Akhmedov, Rubakov and Smirnov (ARS) where sterile neutrinos with masses of a few GeV can generate the BAU through their CP-violating oscillations \cite{Akhmedov:1998qx}. In this scenario, total lepton number is approximately conserved at high temperatures, but it is redistributed among the active and sterile sectors. Lepton number in the active sector is then partially converted into a baryon asymmetry via sphaleron transitions. In this case, one assumes that no RHNs were present right after inflation and the small Yukawa couplings connecting the active and sterile sectors guarantee that at least one of the sterile states remains out-of-equilibrium until sphaleron freeze-out. ARS is therefore a \emph{freeze-in scenario}. 

However, one of the main drawbacks of the vanilla ARS is that it requires significant mass degeneracy among the sterile neutrinos to achieve the observed BAU \cite{Drewes:2017zyw}. Moreover, although lighter sterile neutrinos are kinematically accessible in many current experiments, this advantage is counterbalanced by the tiny Yukawa couplings $\sim \mathcal{O}(10^{-7})$ which are very challenging to probe \cite{Beacham:2019nyx}. To resolve this issue, and further motivated by the potential occurrence of additional fields in nature, some attention has been given to models featuring interactions among the sterile neutrinos (see e.g. \cite{Flood:2021qhq, Fischer:2021nha}). Certainly, new interactions in the sterile sector could improve future detection prospects. Yet, it has been shown that including interactions that could bring the RHNs into equilibrium earlier can lead to a reduction in the generated BAU and, in best-case scenarios, only to an $\mathcal{O}(1)$ enhancement with respect to the vanilla ARS \cite{Flood:2021qhq, Fischer:2021nha}. Other extensions of the minimal sterile neutrino framework and their impact on ARS leptogenesis were studied, e.g. in \cite{Escudero:2021rfi,Caputo:2018zky, Alanne:2018brf}.  

In this work, we focus on the first issue, namely the mass degeneracy in the vanilla ARS. The goal of this work is to investigate if self-interactions among the sterile neutrinos can relax the mass degeneracy required in vanilla ARS. For this purpose we study a  setup with two sterile neutrinos interacting via a scalar singlet. This is similar to the model considered in \cite{Fischer:2021nha,Flood:2021qhq} but we allow for a more general structure of Yukawa couplings than these previous works. 

This paper is structured as follows. Section~\ref{sec: Vanilla ARS} provides a brief introduction to the ARS mechanism, outlining how the asymmetry is generated in the vanilla scenario, and we present the quantum kinetic equation (QKE) governing the evolution of the BAU. In Section~\ref{sec: neutrino self-interactions}, we explore the inclusion of the effects of neutrino self-interactions in the equations and present the corresponding numerical results. To put this in perspective we derive an approximate analytical solution in the oscillatory regime and highlight the key differences between the vanilla ARS and the extended model. Finally, Section~\ref{sec: conclusions} offers our conclusions and provides an outlook for future work.

\section{Basics of freeze-in leptogenesis}
\label{sec: Vanilla ARS}

\subsection{The ARS mechanism}

We begin by giving a brief review of the ARS mechanism \cite{Akhmedov:1998qx} and summarizing how the baryon asymmetry is generated within this scenario (for a more complete discussion see e.g. the review \cite{Drewes:2017zyw}). The Lagrangian of the model corresponds to a type I seesaw mechanism. In its minimal incarnation, the SM is extended by two right-handed neutrinos\footnote{Even though in the original ARS mechanism the SM is extended with three sterile neutrinos, only two are enough to generate a baryon asymmetry. For the sake of simplicity we, therefore, consider this minimal case.}. The Lagrangian thus reads
\begin{equation}
    \mathcal{L} = \mathcal{L}_{\text{SM}} +  i \bar{N}_i  \slashed{\partial} N_i- \frac12 \bar{N}_i M_{ij}N_j - \left(F^{\dagger}_{a i }\bar{L}_{a}\epsilon \Phi N_i +h.c.\right ),
    \label{eq: lagrangian}
\end{equation}
where $\mathcal{L}_{\text{SM}}$ is the Lagrangian of the SM, $N_i$ are the RH neutrinos and $\Phi$ and $L = (\nu_L, e_L)^T$ are the SM Higgs and left-handed lepton doublets, respectively. Here, $M$ is the Majorana mass matrix of the RH neutrinos, $F$ are the Yukawa matrices connecting the sterile sector with the SM and $\epsilon$ is the $2 \times 2$ antisymmetric tensor, i.e. $\epsilon^{12}=-\epsilon^{21}=1.$ From now on we will work in the mass basis, i.e. the basis in which $M$ is diagonal. 

As pointed out by Sakharov in the 70's, a mechanism capable of generating the baryon asymmetry must fulfill three conditions \cite{Sakharov:1967dj}: i) baryon number violation, ii) C and CP violation and iii) deviation from equilibrium. In the SM baryon number and C-symmetry are violated by the weak sphalerons and the weak interactions, respectively. However, CP violation is small in the SM and, in addition, a significant departure from thermal equilibrium is not expected. 
The first of these shortcomings is naturally addressed by the seesaw mechanism since it  provides additional sources of CP violation in the mixing in the neutrino sector, i.e. encoded in complex phases present in the Yukawa coupling matrix $F$. Moreover, new processes involving the sterile neutrinos open up the possibility of departure from thermal equilibrium. For instance, in its original proposal, thermal leptogenesis relies on the out-of-equilibrium decay of the heavy RHNs, whereas in the ARS mechanism non-equilibrium is established due to the tiny couplings of light steriles which are produced from the SM through freeze-in.

 The creation of the baryon asymmetry in the ARS model unfolds via the interplay of multiple interlocking processes. In the oscillatory regime, these neatly separate into three stages that are controlled by different timescales. For convenience, let us define a dimensionless time variable $z = T_{\text{ws}}/T$, where $T_{\text{ws}}$ is a reference temperature that is taken to be $T_{\text{ws}} \simeq 130~ \text{GeV}$, i.e. the temperature at which sphalerons freeze-out. In the first stage, right-handed neutrinos are produced at very high temperatures in flavour states. Since $F$ and $M$ are not diagonal in the same basis, they start oscillating at a scale $z_{\text{osc}}$ given by \begin{equation}
    z_{\text{osc}} = T_{\text{ws}} \left(\frac{216~ \zeta(3)}{a_R~ \pi~ |\Delta M^2|}\right)^{1/3},
\end{equation}
with the squared mass difference defined as $|\Delta M^2| = |M_1^2 - M_2^2|$. Here $a_R = m_{Pl} \sqrt{\frac{45}{4 \pi ^3 g_* }}$ with $m_{Pl} = 1.22 \times 10^{19} \; \text{GeV}$ being the Planck mass and $g_*= 106.75$ denoting the effective number of relativistic degrees of freedom. These CP-violating oscillations lead to the generation of flavour asymmetries in the active sector
\begin{equation}\mathtt{L}_{a}(z > 0) = \frac{n_a - n_{\bar{a}}}{s} \neq 0,   \quad a= e, \mu,\tau, 
\label{eq: sterile lepton number}
\end{equation} with $s$ the entropy density of the universe and $n_{a (\bar{a})}$ the number densities of the active (anti) leptons.

 However, since there is no total lepton number violating processes\footnote{This holds for $z<1$ if the Majorana masses for the sterile neutrinos are very small given that total lepton number violation is a $M^2 / T $ effect \cite{Hambye:2017elz}.}, total lepton number remains zero throughout, namely  \begin{equation}
    \mathtt{L}_{\text{tot}} = \sum_{i =1,2} q_{N_i} + 2 \sum_{a= e, \mu,\tau} \mathtt{L}_a = 0.
    \label{eq: conservation of total lepton number}
\end{equation}
Here the factor of two accounts for the $\text{SU}(2)$ multiplicity of the active doublets and the sterile charges are analogously defined as
\begin{equation}q_{N_{i}} = \frac{n_i^+ - n_i^-}{s},   \quad i= 1,2, 
\label{eq: sterile lepton number}
\end{equation} where we denote the "plus" and "minus" helicities, which play the role of particle and antiparticles in the sterile sector. 
Note that, through the Yukawa couplings, lepton number is redistributed between the sterile and active sectors. 

In the final step, as long as the temperature remains above $T_{\text{ew}}$ (for $z \lesssim 1$), a portion of the active lepton asymmetries $\mathtt{L}_a$ is reprocessed by weak sphalerons, resulting in a net baryon number.
 It is interesting to note, that for $z \ll z_{\text{osc}}$ the oscillation phase is still very small and the generation of the flavour asymmetries is suppressed \cite{Flood:2021qhq}. 
On the other hand, for $z \gg z_{\text{osc}}$ the oscillations become very rapid and the contributions to the asymmetry average to zero. 
The bulk of the asymmetry is therefore produced around $z \sim z_{\text{osc}}$, corresponding to the time when the sterile neutrinos undergo the first oscillations.   
 Another crucial time scale in the system is the point at which the abundance of right-handed neutrinos reaches thermal equilibrium, at $z_{\text{eq}} \propto \left(||F^* F^T ||\right)^{-1}$, with $|| \cdot ||$ denoting the absolute value of the largest eigenvalue. Importantly, at least one neutrino species must remain out of equilibrium until after $z = 1$, as the third Sakharov condition would be violated otherwise. 
 
If the Yukawa couplings are sufficiently small one can get an analytical estimation for the resulting asymmetry by using a perturbative calculation in $F$ \cite{Asaka:2005pn, Hambye:2017elz, Drewes:2016gmt} which leads to
\begin{equation}
    \frac{B}{s} \propto  |F^*F^T|^3 \left(\frac{\Delta M ^2}{\text{GeV}^2}\right)^{-2/3}. 
    \label{eq:ARS asymmetry}
\end{equation}
This result is well-known and it is one of the main drawbacks for the ARS mechanism. Plugging in numbers for the Yukawa couplings $F$, one finds that, one can obtain the observed value of the BAU only for rather degenerate neutrino masses $ |\Delta M ^2|/ M_i M_j \lesssim  10^{-4} ~ \left(10^{-3}\right)$ for normal (inverted) hierarchy for the active neutrino masses \cite{Hernandez:2015wna, Hernandez:2016kel}. Less degenerate spectra are possible if lepton number violation is included  \cite{Eijima:2018qke} or if three RH neutrinos are considered instead of two \cite{Drewes:2012ma}. 

\subsection{The quantum kinetic equation}
In the most general case, the dynamics of the system is governed by the kinetic equation for the complete neutrino density matrix $\rho$. It  reads~\cite{Dolgov:1980cq, Sigl:1993ctk, Barbieri:1990vx} 
\begin{equation}
    i \frac{d \rho}{dt} = \left[\mathbb{H}, \rho\right] - \frac{i}{2} \left\{ \Gamma, \rho \right\} + \frac{i}{2} \left\{ \Gamma^p , \mathbb{I} - \rho\right\},
    \label{eq: Full QKE}
\end{equation}
where $\mathbb{H}(k) = H^0(k) + V(k)$ is the effective Hamiltonian, consisting of the free Hamiltonian $\left[H^0(k)\right]_{ij} =  \left(k^2 + M_I^2\right)^{1/2} \delta_{ij}$ and the effective potential $V(k)$ which is induced by medium effects. Here, $k$ represents the momentum of the neutrinos, $M_I$ their corresponding masses, while $\Gamma$ and $\Gamma^p$ are the destruction and production rates, respectively. Following Refs.~\cite{Asaka:2005pn, Asaka:2011wq} we can approximate the last term in~\eqref{eq: Full QKE} simply as $\Gamma^p$ assuming Boltzmann statistics, i.e. $\mathbb{I} - \rho \approx \mathbb{I}$. 
Moreover, since the equilibrium density matrix $\rho_{\text{eq}}=\exp (- \mathbb{H}/T)\approx \exp(-k/T)$
must satisfy~\eqref{eq: Full QKE} one can rewrite the production rate as $\Gamma^p = \frac{1}{2} \left\{ \Gamma, \rho_{\text{eq}}\right\}.$ 
Formally, $\rho$ is a $10 \times 10$ matrix, where the diagonal elements correspond to occupation numbers whereas the off-diagonal elements describe flavour correlations. By performing a series of simplifications (see e.g.~\cite{Asaka:2005pn}) $\rho$ is reduced to a block-diagonal matrix, consisting of   two $3 \times 3$ matrices describing the active (anti-)neutrino system and two $2 \times 2$ non-diagonal matrices representing the sterile sector. 
Since the oscillations of the active neutrinos do not play a role on the scales considered here, their sub-matrices can be reduced to the diagonal elements corresponding to the occupation numbers. The sterile neutrinos, on the other hand, need to be treated 
using a full quantum approach, such that 
all the elements of their density matrix remain relevant.
It follows that the dynamics of the system will be determined by the evolution of the eigenvalues of $\mathbb{H}$ and $\Gamma$. In fact, it is precisely the misalignment between $\mathbb{H}$ and $\Gamma$, which in general are not diagonal in the same basis, that leads to the oscillations in the sterile sector. After the simplifications, one is left with two $2\times2$ density matrices for the sterile neutrinos and six active (anti-)neutrino occupation numbers.

For simplicity, from now on, we will work with an integrated version of Eq.~\eqref{eq: Full QKE} that allows us to remove the momentum dependence from both sides of the evolution equation. 
With this in mind we start by parametrizing the density matrix for the sterile neutrinos as 
\begin{equation}
    \rho_{N^{\pm}} (k) = \rho_{\text{eq}}(k) + \delta \rho_{N^{\pm}}, 
    \
\end{equation}
 where we indicate the "plus" and "minus" helicities and $\delta \rho$ denotes the deviation from equilibrium which we assume is independent of momenta. Here and in the following we adopt the notation of \cite{Drewes:2016gmt} and use their expressions. Further one can take the thermal average of all the operators, i.e. 
\begin{equation}
    \left<  \mathcal{O}\right> = \frac{\int dk k^2 \mathcal{O} \rho_{\text{eq}}}{\int dk k^2 \rho_{\text{eq}}},
    \label{eq: thermal average}
\end{equation}
with $\rho_{\text{eq}}$ the equilibrium distribution. Additionally, it is convenient to rewrite Eq.~\eqref{eq: Full QKE} in terms of the dimensionless quantity $z$. To this end, we use the time derivative in an expanding universe given by 
\begin{equation}
    \frac{d}{dt} = \frac{\partial}{\partial t} - H k \frac{\partial}{\partial k},
\end{equation}
where $H = \frac{T^2}{a_R}$ is the Hubble expansion rate for a radiation dominated universe. By incorporating all these simplifications, we obtain an equation describing the evolution of the deviations of the heavy neutrinos' number density from equilibrium. Defining $\delta n$ in analogy with $\delta \rho$ one finds \cite{Drewes:2016gmt}
\begin{equation}
   \frac{d\delta n_{N^{\pm}}}{dz} = -\frac{i}{2} \left[ H_{N^{\pm}}^{\text{th}} + z^2 H_{N^{\pm}}^{\text{vac}}, \delta n_{N^{\pm}}\right] - \frac12 \left\{\Gamma_{N^{\pm}}, \delta n_{N^{\pm}}\right\} + \sum_{a,b = e, \mu, \tau} \tilde{\Gamma}^a_{N^{\pm}} \left( A_{ab} + C_b/2\right)\Delta_b.
   \label{eq: integrated Boltzmann eq}
\end{equation}
Here, the first term on the right-hand side contains the contributions from the Hamiltonian, the second term accounts for the contributions from the rate, and the third term corresponds to the SM back reaction. The effective Hamiltonian in vacuum $H_{N^{\pm}}^{\text{vac}}$ and the corresponding finite temperature corrections $H_{N^{\pm}}^{\text{th}}$ are given by 
\begin{align}
    H_{N^{\pm}}^{\text{vac}} &= \frac{\pi^2}{18 \zeta(3)} \frac{a_R}{T_{\text{ws}}^3} \left( \text{Re} [M^{\dagger} M] \pm i \text{Im}[M^{\dagger}M]\right), \nonumber \\ 
    H_{N^{\pm}}^{\text{th}} & = h_{\text{th}} \frac{a_R}{T_{\text{ws}}} \left( \text{Re}[F^* F^T] \mp i \text{Im}[F^* F^T]\right),
\end{align}
with $h_{\text{th}} \approx 0.23$. Meanwhile, the rates appearing in the collision terms of Eq.~\eqref{eq: integrated Boltzmann eq} are defined as follows
\begin{align}
    \Gamma_{N^{\pm}} &= \gamma_{AV} \frac{a_R}{T_{\text{ws}}} \left( \text{Re}[F^* F^T] \mp i \text{Im}[F^*F^T] \right),\nonumber \\ 
    \left(\tilde{\Gamma}^a_{N^{\pm}}\right)_{ij} &= \pm \frac12 \gamma_{AV} \frac{a_R}{T_{\text{ws}}} \left(\text{Re}[F^*_{ia} F^T_{aj}] \mp \text{Im}[F^*_{ia} F^T_{aj}] \right),
\end{align}
where $\gamma_{AV}= 0.012.$ The coefficients on the last term of Eq.~\eqref{eq: integrated Boltzmann eq} are given by 
\begin{equation}
    A= \frac{1}{711}\begin{pmatrix}
    -221 & 16 &16\\ 
    16 & -221 & 16 \\ 
    16 &16 & -221
    \end{pmatrix}, \quad C = -\frac{8}{79} \begin{pmatrix}
        1&1&1
    \end{pmatrix}.
\end{equation}
On the other hand, the evolution of the SM-conserved charges  $\Delta_a = B/3 - L_a$ is described by the equation
\begin{equation}
    \frac{d \Delta_a}{dz} = \frac{\gamma_{AV}}{2} \frac{a_R}{T_{\text{ws}}} \sum_i F_{ia} F^{\dagger}_{ai} \left( \sum_{b} (A_{ab}+ C_b/2) \Delta_b - q_{N_i}\right) - \frac{S_a}{T_{\text{ws}}},
    \label{eq: full evolution of flavour charges}
\end{equation}
with the source term 
\begin{equation}
    S_a = \gamma_{AV} a_R \sum_{\substack{i,j \\ i\neq j}} F^*_{ia} F_{ja} \left\{i \text{Im} [(\delta n_N)_{ij}^{\text{even}} ] + \text{Re}[(\delta n_{N})_{ij}^{\text{odd}}]\right\}\,.
    \label{eq: source term}
\end{equation}
Here we have defined the even and odd contributions as 
\begin{equation}
    \delta n_N^{\text{even}} = \frac{\delta n_{N^+} + \delta n_{N^-}}{2} \quad \text{and}\quad \delta n_{N}^{\text{odd}} = \frac{\delta n_{N^+} - \delta n_{N^-}}{2},
\end{equation}
so that the sterile charges are given by $q_{N_i} = 2 (\delta n_{N})_{ii}^{\text{odd}}.$
It is important to note that Eq.~\eqref{eq: integrated Boltzmann eq} is valid only for relativistic neutrinos, as we neglect their mass in all terms except for $H_{N^{\pm}}^{\text{vac}}$. This approximation is reasonable provided that the masses of the heavy neutrinos are of the order of a few GeV. The net baryon asymmetry is frozen-in at the time of sphaleron freeze-out and is given by 
\begin{equation}
    B = \frac{28}{79} \left[ \left.\Delta_1(z) + \Delta_2(z) + \Delta_3(z)\right]\right|_{z=1} . 
\end{equation}

A detailed review of the analytical solution of Eqs.~\eqref{eq: integrated Boltzmann eq} and \eqref{eq: full evolution of flavour charges} using a perturbative expansion in $F$ \cite{Drewes:2016gmt} is provided in Appendix~\ref{sec: Garbrecht analytical approach}.

\section{Sterile neutrino self-interactions}
\label{sec: neutrino self-interactions}

In this section we explore how the picture changes when we allow for self-interactions among the sterile neutrinos. Specifically, we will consider interactions of the neutrinos with a scalar singlet $\phi$. For simplicity, we assume $\phi$ to be in thermal equilibrium with the SM plasma. The interaction Lagrangian for the scalar-sterile neutrino coupling is given by
\begin{equation}
    \mathcal{L}_{\text{int}} = Y_{ij}~ \bar{N}_i N_j \phi,
    \label{eq: interaction term}
\end{equation}
where $Y$ is a $2 \times 2$ real and symmetric matrix. This new term would lead to the generation of masses for the sterile neutrinos if $\phi$ acquires a vacuum expectation value. If this is the only contribution to the mass, $Y$ is diagonal in the same basis as the mass matrix $M$.  However, in our discussion, we do not assume a specific origin for the heavy neutrino masses, allowing for the more general case where $Y$ and $M$ are independent. This is the key difference between this study and previous works of ARS leptogenesis in scalar extensions where, either a diagonal Yukawa matrix was assumed~\cite{Flood:2021qhq} or the neutrino masses where generated by $\phi$ directly \cite{Fischer:2021nha}. As we will see later, the fact that we allow $M$ and $Y$ to be diagonal in different bases leads to an enhancement of the produced BAU through thermal effects in the dispersion relation of the sterile neutrinos. 
\begin{figure}[t!]
\centering
\begin{subfigure}{0.4\textwidth}
\centering
\raisebox{25pt}{\begin{tikzpicture}
\begin{feynman}
\vertex (a1) {\(N_i\)};
\vertex[right=1.5cm of a1] (a2);
\vertex[right=3cm of a2] (a3);
\vertex[right=1.0cm of a3] (a4) {\(N_i\)};

\diagram* {
{[edges=fermion]
(a1) -- [] (a2) -- [edge label=\(N_j\), line width = 0.7mm] (a3) --  (a4),
},
(a2) -- [scalar,out=90, in=90, looseness=1.5, edge label=\(\phi\), line width = 0.7mm] (a3),

};
\end{feynman}
\end{tikzpicture}}
\end{subfigure}%
\hspace{1cm}
    ~ 
\begin{subfigure}{0.4 \textwidth}
\centering
    \begin{tikzpicture}
\begin{feynman}
\vertex (a3);
\vertex[right=1.5cm of a3] (a4);
\vertex[right = 1.5 cm of a4](a2);
\vertex[above = 1 cm of a2] (a5){\(N_i\)};
\vertex[below = 1 cm of a2] (a6){\(N_j\)};
\diagram* {
(a3) -- [scalar, edge label=\(\phi\)] (a4),
(a4) -- [fermion] (a5), 
(a6)-- [fermion] (a4)
};
\end{feynman}
\end{tikzpicture}
\end{subfigure}
\caption{Leading order contributions from the new interaction of RH neutrinos in freeze-in leptogenesis.}
\label{fig: Feynmann diagrams}
\end{figure}
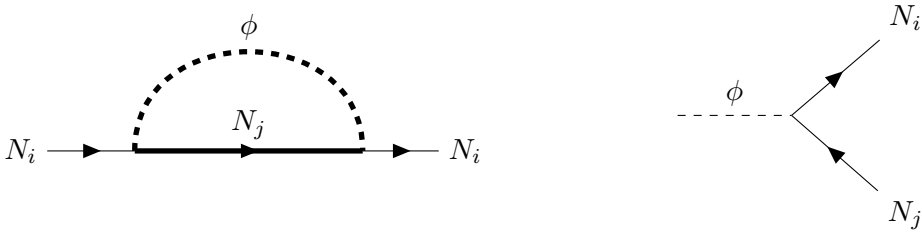

The interaction term modifies Eq.~\eqref{eq: Full QKE} in two ways. First, forward scattering of sterile neutrinos in the hot plasma alters their dispersion relation, appearing as a new contribution to the thermal potential \cite{Notzold:1987ik,Quimbay:1995jn}, i.e. $\mathbb{H} = H^0 + H^{\text{th}}_{\text{SM}} + H^{\text{th}}_{\phi}$. The lowest order corrections appear at one loop level in the bubble diagram shown in the left panel of Fig.~\ref{fig: Feynmann diagrams}. Assuming the scalar to be sufficiently light ($m_{\phi} \ll T$), the thermal potential takes the form
\begin{equation}
   H^{\text{th}}_{\phi}(k) = 
        Y \cdot Y^T \, \frac{  T^2 }{24 k } \rightarrow \left<H^{\text{th}}_{\phi}\right> =  Y \cdot Y^T \, \frac{  T \pi^2}{432 \zeta(3)}, 
        \label{eq: sterile potential}
\end{equation}
where we have taken the thermal average as in Eq.~\eqref{eq: thermal average}, see e.g.~\cite{Astros:2023xhe} for more details on the computation of the potential. Second, the new interaction introduces additional production and destruction channels for the sterile neutrinos via the decay and inverse decay of the scalar (see right panel of Fig.~\ref{fig: Feynmann diagrams}). 
For the decay of the scalar $\phi \rightarrow \bar{N}_i N_j$ the thermally averaged rate is given by
\begin{equation}
    \left<\Gamma^p_{\phi}\right> \approx Y \cdot Y^T \frac{m_{\phi}^2 T^2}{72 \pi \, n_{\phi}^{\text{eq}}},
\end{equation}
where $n_{\phi}^{\text{eq}}$ is the equilibrium number density of the scalar. 
With these two new contributions Eq.~\eqref{eq: Full QKE} takes the following form
\begin{equation}
     i \frac{d \rho}{dt} = \mathbb{C}_{\text{ARS}} + \left[H^{\text{th}}_{\phi}, \rho\right] - i \rho ~\Gamma^{d}_{\phi}~ \rho + i \Gamma^{p}_{\phi},
     \label{eq: Full QKE new physics}
\end{equation}
where for the sake of readability we denote as $\mathbb{C}_{\text{ARS}}$ the contribution from the ARS mechanism described in the previous section (see Appendix \ref{app: derivation of the QKE} for a careful derivation of Eq.~\eqref{eq: Full QKE new physics}). Note that for the destruction rate associated with the new interaction, the second power of the density  $\rho$ is appears, as the inverse decay of the scalar involves the annihilation of two sterile neutrinos. The destruction rate is connected to the production rate through the relation 
\begin{equation}
    \Gamma^{d}_{\phi} = \frac{\Gamma^p_{\phi}}{\rho_{\text{eq}}^2}. 
\end{equation}
As we will see, the new physics back reaction rate is not relevant in the parameter space we discuss in the following sections. The integrated version of Eq.~\eqref{eq: Full QKE new physics} is 
\begin{align}
     \frac{d\delta n_{N^{\pm}}}{dz} =& -\frac{i}{2} \left[ H_{N^{\pm}}^{\text{th}} + z^2 H_{N^{\pm}}^{\text{vac}} , \delta n_{N^{\pm}}\right] - \frac12 \left\{\Gamma_{N^{\pm}}, \delta n_{N^{\pm}}\right\} + \sum_{a,b = e, \mu, \tau} \tilde{\Gamma}^a_{N^{\pm}} \left( A_{ab} + C_b/2\right)\Delta_b \nonumber \\ &-\frac{i}{2} \left[  V_{\phi} , \delta n_{N^{\pm}}\right] - \frac{z^2}{n_{\text{eq}}} \left\{\Gamma_{\phi}, \delta n_{N^{\pm}}\right\} - \frac{z^2}{ \, n^2_{\text{eq}}}  \delta n_{N^{\pm}} ~\Gamma_{\phi}~ \delta n_{N^{\pm}},
   \label{eq: integrated Boltzmann Eq new physics}
\end{align}
where, for simplicity, we have written all the contributions from the new interactions in the second line and the additional terms are given by 
\begin{align}
    V_{\phi} & = \frac{\pi^2 a_R}{432 \, \zeta(3) T_{\text{ws}}} \, Y \cdot Y^T, \qquad 
    \Gamma_{\phi} = \frac{\pi m_{\phi}^2 \, a_R\,  n^{\text{eq}}_{\phi}}{72 \zeta(3)  \, T_{\text{ws}}^3} \, Y \cdot Y^T.   
\end{align}

\subsection{Numerical solutions}

\begin{table}[t!]
    \centering
    \begin{tabular}{|c|c|}
        \hline
       \multicolumn{2}{|c|}{Parameters}   \\
        \hline
       $m_2^2 = 7.41 \times 10^{-5}~ \text{eV}^2$ & $M_{1} \approx M_{2} = 1 ~ \text{GeV}$ \\ 
         
        $m_3^2 = 2.511 \times 10^{-3} ~ \text{eV}^2$ &$\delta = \frac{3 \pi}{2}$
          \\ 
         $\sin^2(\theta_{12})=0.307 $   &$\alpha_1 = 0 $
\\ 
 $\sin^2( \theta_{13}) = 0.02203$ & $\alpha_2 = -2\pi$ \\ 
 $\sin^2(\theta_{23}) = 0.572$ & $\omega = \frac{3\pi}{4} + 2.16 i$ \\
 \hline 
 
    \end{tabular}
    \caption{The parameters used for the Casas-Ibarra parametrization of $F$.}
    \label{tab: parameters for the Yukawa coupling}
\end{table}

In this section we will investigate the numerical solutions to Eq.~\eqref{eq: integrated Boltzmann Eq new physics}. We will restrict ourselves to the oscillatory regime allowing us to make the connection to the discussion in Appendix~\ref{sec: Garbrecht analytical approach}. Therefore we choose a rather small Yukawa coupling $F$ that we keep fixed for the rest of this work. For this we use the Casas-Ibarra parametrization \cite{Casas:2001sr} (see Appendix~\ref{app: Casa-Ibarra parametrization}) with the parameters specified in Tab.~\ref{tab: parameters for the Yukawa coupling}. 
The parameters in the left column of the table, are taken from global analyses of different neutrino oscillation experiments \cite{Gonzalez-Garcia:2021dve} assuming the lightest of the active neutrinos to be massless and a normal mass ordering. The parameters in the right column cannot be directly constrained by oscillation experiments and we take the benchmark values used in \cite{Drewes:2016gmt} for convenience. In addition, for simplicity, we will consider the special case  $Y_{ii}=Y_{jj}$. We fix the mass of the scalar to $m_{\phi} = 3~\text{GeV}$ throughout.

\begin{figure}
    \centering
\includegraphics[width=1.05\linewidth]{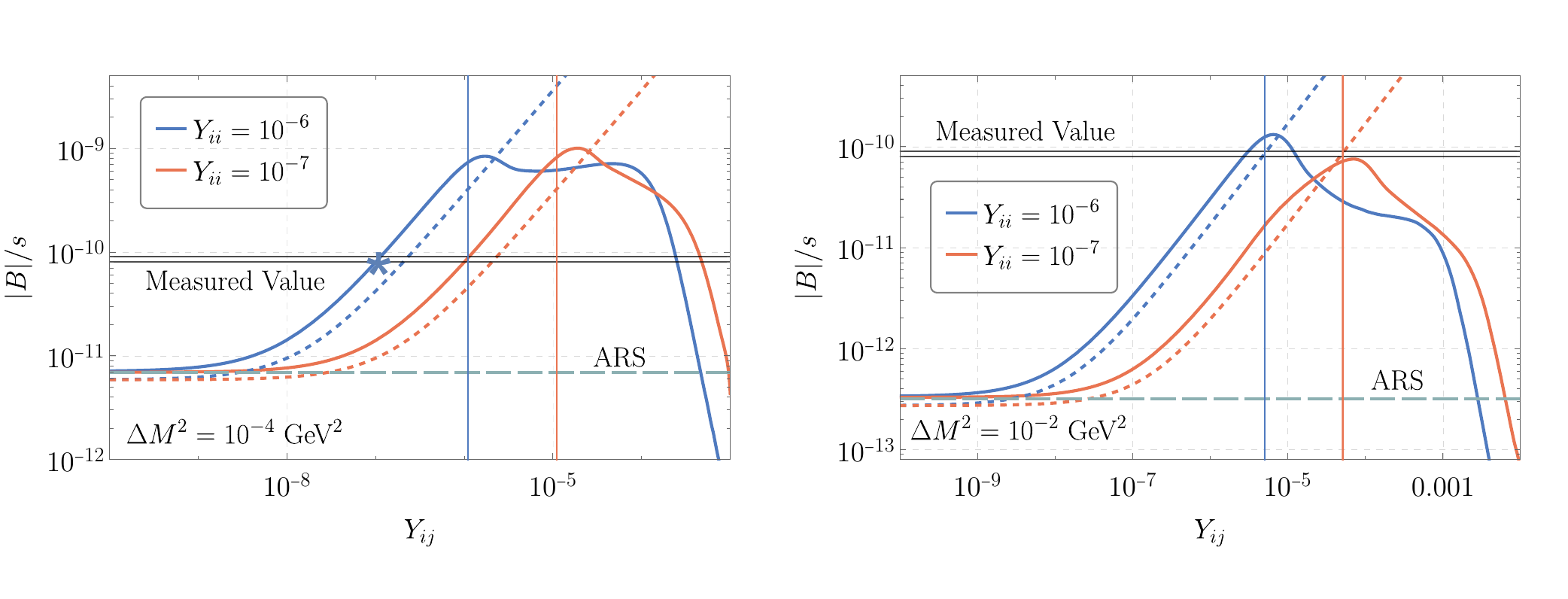}
    \caption{The baryon asymmetry with respect to the off-diagonal Yukawa components for a fixed value of the mass of the mediator $m_{\phi} = 3~\text{GeV}$ and two different values for the diagonal components $Y_{ii} = 10^{-6}$ (solid blue) and $Y_{ii} = 10^{-7}$ (solid orange) and two different mass splittings $\Delta M^2 = 10^{-4} $ (left) and $\Delta M^2 = 10^{-2} $ (right). In light blue we indicate the value of the asymmetry obtained with the vanilla ARS for the corresponding mass splittings and in black the measured BAU. For completeness, as dashed lines we also show the analytical approximation (see Sec.~\ref{sec: Analytical solution} for more details) and the values of the Yukawa for which we expect it to break down. The blue star (left) indicates the benchmark discussed in the text.}  
    \label{fig: asymmetry vs Yukawa}
\end{figure} 
\begin{table}[]
    \centering
    \begin{tabular}{|c|c|c|c|c|}
        \hline
       \multicolumn{5}{|c|}{Benchmark point}   \\
        \hline
         $\Delta M ^2$ & $Y_{ii}$ & $Y_{ij}$ & $m_{\phi}$ & $ |B|/s$\\
        \hline
       $10^{-4} ~ \text{GeV}^2$ & $10^{-6}$& $1.1 \times 10^{-7}$ & $3 ~\text{GeV}$ & $ 8.76 \times 10^{-11}$ \\ 
 \hline 
 
    \end{tabular}
    \caption{Parameters corresponding to the benchmark point discussed in the main text.}
    \label{tab: benchmark point}
\end{table}
In Figure~\ref{fig: asymmetry vs Yukawa} we show the evolution of the asymmetry as a function of the off-diagonal components of the Yukawa coupling $Y_{ij}$ for two fixed values of $Y_{ii} = 10^{-6}$ (blue) and $Y_{ii} = 10^{-7}$ (orange). The left panel corresponds to a mass splitting of $\Delta M^2 = 10^{-4}~\text{GeV}^2$ while the right panel corresponds to $\Delta M^2 = 10^{-2} ~\text{GeV}^2$. For comparison, we also indicate with a dashed light blue line the value of the BAU that is generated in the vanilla ARS scenario for each mass splitting. The two solid black lines denote the range in which the measured value of the BAU lies, i.e. between $8 \times 10^{-11}$ and $9 \times 10^{-11}.$ In addition, the dashed blue and orange lines correspond to the analytical solutions discussed in detail in Sec.~\ref{sec: Analytical solution}, while the vertical solid lines mark the values of the off-diagonal Yukawa couplings where the analytical approximations are expected to break down. 
We see that by the inclusion of the new interactions the generated BAU is enhanced by over two orders of magnitude. As a result, the measured value of the BAU can be achieved for square mass splittings significantly larger than those in the vanilla ARS case.  
As we will discuss in detail in the next section, the increase in the BAU observed to the left of the vertical lines is primarily driven by the effects of the sterile thermal potential. However, when the thermal potential becomes too large compared to $H_{N^{\pm}}^{\text{vac}}$ 
the asymmetry ceases to grow and begins to decline. The reason for this is essentially the same behind the suppression of the asymmetry by $(\Delta M^2)^{-2/3}$ in the vanilla ARS. In that case, a larger mass splitting results in a smaller typical oscillation scale $z_{\text{osc}}$, i.e. oscillations start at earlier times and there is a smaller population of sterile neutrinos at that moment which leads to a suppression. This is also true in a system where the oscillations are dominated by the squared thermal mass difference. In analogy to the vacuum mass, the thermal mass is defined as the eigenvalues of the thermal potential $V_N$ such that the thermal mass difference is 
\begin{equation}
    \Delta M _{\text{thermal}} =  \frac{4~ \pi^2 a_R}{432 \zeta(3) T_{\text{ws}}}   Y_{ii}Y_{ij}.
\end{equation}
In a similar way, in the limit $\Delta M^2 \rightarrow 0$ the asymmetry is suppressed by $\Delta M_{\text{thermal}}^{-1}$. In the regime to the right of the solid vertical lines, the interplay between the two different oscillation scales makes the problem non-trivial and the suppression of the asymmetry by the thermal potential follows a different scaling.
Increasing the coupling further, we note a subtle rise in the generated asymmetry. We assume that is due to the rate of decay of the scalar into two RHNs becoming effective but we do not have an analytic result that covers this regime. In any case, this behavior is transitional. At even larger coupling, the rate equilibrates the sterile neutrinos more efficiently than the SM one such that  equilibrium is reached before sphaleron freeze-out. This leads to the strong suppression of the BAU that is apparent to the right side of Fig. ~\ref{fig: asymmetry vs Yukawa}. This decrease is in agreement with the results of \cite{Fischer:2021nha,Flood:2021qhq} whereas the increase, that depends on $Y_{ij}$, has been missed by them.

\begin{figure}
    \centering
    \includegraphics[width=0.7\linewidth]{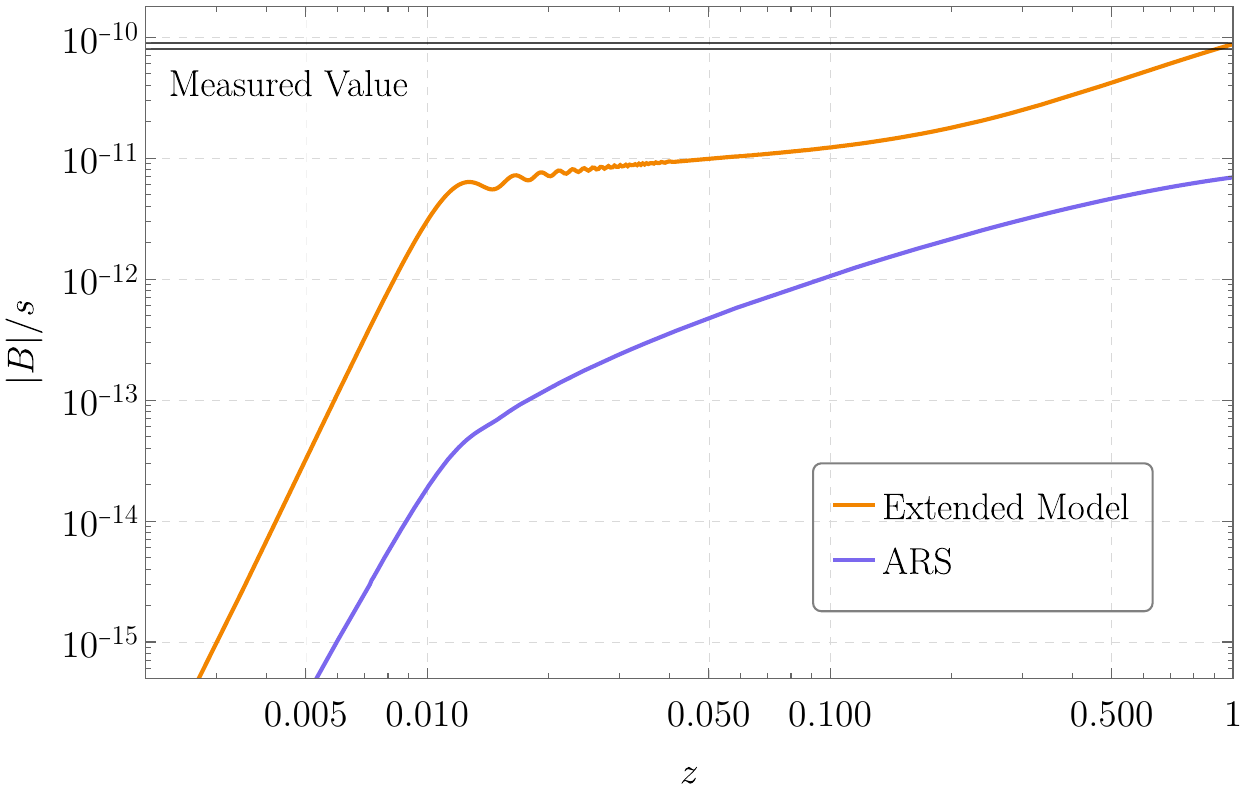}
    \caption{The evolution of the baryon asymmetry with respect to $z$ for the benchmark point considered in this section (solid orange) and the vanilla ARS (solid purple) for comparison. The black horizontal lines denote the measured value of $B$.}
    \label{fig: baryon asymmetry vs z}
\end{figure}

\begin{figure}
    \centering
    \includegraphics[width=0.7\linewidth]{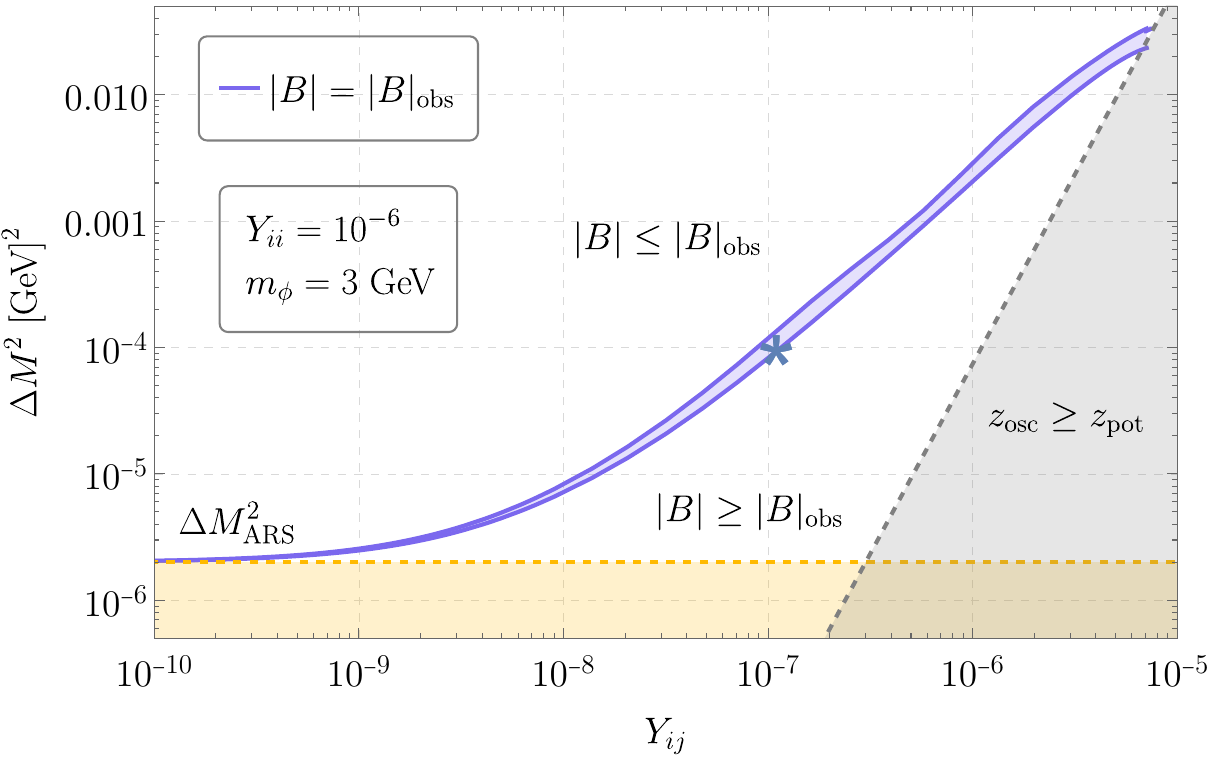}
    \caption{The parameter space in the $Y_{ij} - \Delta M^2$ plane for fixed values of $Y_{ii} = 10^{-6}$ and $m_{\phi} = 3 ~\text{GeV}$. Everywhere between the purple lines the measured value of the BAU can be obtained. Above the upper purple line, the BAU is too small, while below the lower purple line, the BAU exceeds the observed limits. The dashed yellow line represents the mass splitting required in the vanilla ARS mechanism to achieve $B_{\text{obs}}$ while the gray region indicates the region where our analytical approximation breaks down. For completeness we also indicate  the benchmark point shown in the previous figure with a blue star.}
    \label{fig:parameter space}
\end{figure}

In Figure~\ref{fig: baryon asymmetry vs z} we present the evolution of the BAU as a function of $z$ for the benchmark point marked by a star in the left panel of Fig.~\ref{fig: asymmetry vs Yukawa}. For this set of parameters, we find an asymmetry consistent with the measured value (see Tab.~\ref{tab: benchmark point}). For comparison, we also include the evolution of the BAU in the vanilla ARS, shown as a solid purple line. With the same mass splitting, the vanilla ARS produces a BAU that falls short by at least one order of magnitude at the time of sphaleron freeze-out. Additionally, we observe that the asymmetry deviates from the vanilla scenario from very early times, with noticeable wiggles appearing around $z = 0.01$. As we will discuss in the next section, these wiggles correspond to oscillations in the sterile charges.

Finally, figure~\ref{fig:parameter space} shows the parameter space in the $Y_{ij}-\Delta M^2 $ plane for a fixed value of $Y_{ii} = 10^{-6}.$ The region between the purple lines can account for the BAU within the observational error. Above the upper line the generated asymmetry is too small, while below the bottom line the BAU exceeds the observational limits. The dashed yellow line indicates the mass splitting required in the vanilla ARS scenario in order to generate $B_{\text{obs}}$. The gray area shows the region for which we expect our analytical approximation to break down. 

\subsection{Analytic solution}
\label{sec: Analytical solution}

In this section we discuss an analytical solution for the scalar extension of the ARS mechanism. For convenience, the solution and its derivation for the standard ARS case are summarized in Appendix~\ref{sec: Garbrecht analytical approach}. To simplify the problem, we assume $Y_{ii} = Y_{jj}$. As noted in section \ref{sec: Vanilla ARS}, the vanilla scenario involves two key time scales: $z_{\text{osc}}$, corresponding to the sterile oscillation phase during which most of the asymmetry is produced, and 
$z_{\text{eq}}$, associated with the equilibration of the neutrinos, which results in the washout of the asymmetry. In this work, we focus on scenarios where there is a distinct separation between the relevant scales, allowing the two processes to be treated independently. In particular, we will focus on the so-called oscillatory regime characterized by $z_{\text{osc}} \ll z_{\text{eq}}$, which occurs, for instance, when the Yukawa couplings are small. In the vanilla ARS scenario, the interplay between these two time scales will determine the evolution of the system.

This remains true for the scalar extension explored in this work. However, now two additional time scales come into play: one associated with the sterile thermal potential $z_{\text{pot}}$ and another linked to the new rate $z_{\text{decay}}$. These are given by 
\begin{equation}
    z_{\text{pot}} = \frac{216 \zeta(3) T_{\text{ws}}}{\pi a_R Y_{ii} Y_{ij}} \quad \text{and} \quad  z_{\text{decay}} = \left(\frac{18 \zeta(3) T_{\text{ws}}^3}{ \pi m_{\phi}^2 a_R Y_{ii} Y_{ij}}\right)^{1/3},
\end{equation}
where $z_{\text{pot}}$ is defined in a manner analogous to $z_{\text{osc}}$, representing approximately the time when the sterile neutrinos undergo a full oscillation driven by the thermal mass difference. Similarly, $z_{\text{decay}}$ denotes the time scale for the decay rate of the scalar to equilibrate. 
For the rate and the potential to play a significant role in the problem, we require $z_{\text{pot, decay}} \leq 1$. Under this condition, we find that the effects of the potential become relevant for $Y_{ij} \cdot Y_{ii}\gtrsim 10^{-14} $, while the rate only becomes important for $Y_{ij}\cdot Y_{ii} \gtrsim 2 \times 10^{-12} $ if we take $m_{\phi} = 3~\text{GeV}$. This allows us to neglect the effects of the rate since they appear at higher couplings and the thermal potential effectively dominates the evolution of the system in the parameter space of interest here.  We then proceed in a similar way to App.~\ref{sec: Garbrecht analytical approach}, performing a perturbative expansion in $F$.  
For $z\ll z_{\text{eq}}$, and neglecting the terms associated with the new rate, we can replace $(\delta n_{N^{\pm}})_{ij} \rightarrow - n_{\text{eq}} \delta_{ij}$ on the right-hand side of Eq.~\eqref{eq: integrated Boltzmann Eq new physics} which leads to
\begin{subequations}
\begin{align}
    \frac{d}{dz} \delta n_{N_{ij}}^{\text{odd}}  + i \Omega_{ij} z^2 \delta n_{N_{ij}}^{\text{odd}} =& \frac{i}{2} V_{\phi_{ij}} \left(\delta n_{N_{ii}}^{\text{odd}} - \delta n_{N_{jj}}^{\text{odd}} \right)  -i \text{Im}[F^* F^T]_{ij} G , \label{eq: full odd off-diagonal} \\  \frac{d}{dz} \delta n_{N_{ij}}^{\text{even}} +i \Omega_{ij} z^2 \delta n_{N_{ij}}^{\text{even}} =& \frac{i}{2} V_{\phi_{ij}} \left( \delta n_{N_{ii}}^{\text{even}} - \delta n_{N_{jj}}^{\text{even}} \right) +  \text{Re}[F^* F^T]_{ij} G, \label{eq: full even off-diagonal}  
\end{align}
\label{eq: Full analytical equations}
\end{subequations}
where $\Omega_{ij}$ and $G$ are defined in Eq.~\eqref{eq: definition of omega and G}. Note that these are the analogues to Eqs.~\eqref{eq: Garbrecht eq}, where now an additional term mixes the off-diagonal and diagonal components through the sterile potential term. To make things simpler we focus on the case for which $z_{\text{osc}} < z_{\text{pot}} \ll z_{\text{eq}}$. This enables us to drop the terms proportional to the potential in Eqs.~\eqref{eq: Full analytical equations} and solve them as in the vanilla case (see Eq.~\eqref{eq: zeroth- order solution}). 
One crucial modification, however, appears in the evolution of the sterile charges, with an additional source term proportional to the potential 
\begin{equation}
    \frac{d q_{N_i}}{dz} = 2 \cdot \mathbb{A}_{\text{ARS}} + 2 V_{\phi_{ij}} \text{Im}[ \delta_{N_{ij}}^{\text{odd}} ], 
\end{equation}
with $\mathbb{A}_{\text{ARS}}$ denoting the right-hand side of Eq.~\eqref{eq: zeroth order sterile charges}. As described in App.~\ref{sec: Garbrecht analytical approach} the vanilla ARS contribution $\mathbb{A}_{\text{ARS}}$ vanishes, so in this scenario in the absence of other sources, no net sterile charges are generated until the SM back reaction becomes relevant. In our case, however, the odd off-diagonal components, through the potential, act as sources for $q_{N_i}$. 

Using the zero-th order solutions in Eq.~\eqref{eq: zeroth- order solution} one can solve for the charges obtaining 
\begin{equation}
    q^{\text{analytical}}_{N_i} (z) = V_{\phi_{ij}}\text{Im}[F^*F^T]_{ij} G \text{Im}\left[i z^2 \, {_2}F_2 \left[\left\{ \frac23 , 1 \right\} ; \left\{ \frac43, \frac53\right\}; -\frac13 i z^3 \Omega_{ij}\right]\right],
    \label{eq: analytical solution for the charges}
\end{equation}
in terms of the generalized hypergeometric function. This result is crucial as it shows that the sterile potential is able to generate charges already at order $\mathcal{O}(F^2)$. This, in turn, implies that a net baryon asymmetry can be produced at order $\mathcal{O}(F^4)$ in contrast to the $\mathcal{O}(F^6)$ suppression in the vanilla ARS scenario. Essentially, two powers of $F$ are replaced by two powers of the new Yukawa coupling $Y$. Once $z \sim z_{\text{eq}}\gg z_{\text{osc}}$ the oscillations of the off-diagonal correlations are too fast and they average to zero. So the evolution of the flavour asymmetries and the sterile charges at this later stage can be described by
\begin{subequations}
\begin{align}
   \frac{d \Delta_a}{dz} &= \frac{\gamma_{AV}}{2} \frac{a_R}{T_{\text{ws}}} \sum_i F_{ia} F^{\dagger}_{ai} \left( \sum_{b} (A_{ab}+ C_b/2) \Delta_b - q_{N_i}\right)
   \label{eq: late wahsout evolution of flavour chrgaes}\\ 
   \frac{d q_{N_i}}{dz} &= - \gamma_{AV} \frac{a_R}{T_{\text{ws}}} \sum_{a} F_{ia} F^{\dagger}_{ai} \left(q_{N_i} - \sum_{b}(A_{ab}+ C_b/2) \Delta_b\right)
\end{align}
\label{eq: late time evolution}
\end{subequations}
just as in the standard case.
Note that Eq.~\eqref{eq: late wahsout evolution of flavour chrgaes} is simply Eq.~\eqref{eq: full evolution of flavour charges} where  the source term has been dropped. The solution to this system of equations is formally given by \cite{Drewes:2016gmt}
\begin{equation}
    \begin{pmatrix}
        \Delta_a (z) \\
        q_{N_i}(z)
    \end{pmatrix} = T \exp\left(\gamma_{AV }\frac{a_R}{T_{\text{ws} }} K_{\text{diag}} z \right) T^{-1} \begin{pmatrix}
        \Delta_a^{\text{osc}}(z_{in}) \\ q_{N_i}^{\text{analytical}}(z_{in})
    \end{pmatrix}; 
    \label{eq: general matrix solution}
\end{equation}
where $z_{in} \gg z_{\text{osc}}$, so that $q_{N_i}^{\text{analytical}}(z_{in})$ and  $\Delta_a^{\text{osc}}(z_{in})$ can be evaluated from Eqs.\eqref{eq: analytical solution for the charges} and ~\eqref{eq: Garbrect initial value flavour charges} taking the limit $z_{in} \rightarrow \infty$. The matrices $K_{diag}$ and $T$ contain the eigenvalues and eigenvectors of a $5 \times 5 $ matrix given by 
\begin{equation}
    K = \begin{pmatrix}
         \frac{1}{2} \sum_{k=1}^{2} F^{\dagger}_{ak} F_{ka} \left(A_{ab} + \frac12\right) && -\frac{1}{2} F^{\dagger}_{aj} F_{ja} \\ 
         \sum_{d=1}^3 F_{id}F^{\dagger}_{di} \left(A_{db} + \frac12 C_b\right) && -\sum_{d=1}^3 F_{id} F^{\dagger}_{di} \delta_{ij}
    \end{pmatrix}
\end{equation}
and $K_{\text{diag}} = T^{-1} K T$. Finally, the baryon asymmetry is given by a simple linear combination 
\begin{equation}
    B = \frac{28}{79} \left[ \Delta_1 (1) + \Delta_2 (1) +  \Delta_3(1)\right].
    \label{eq: analytical B}
\end{equation} 
\begin{figure}{}
    \centering
    \includegraphics[width=0.7\linewidth]{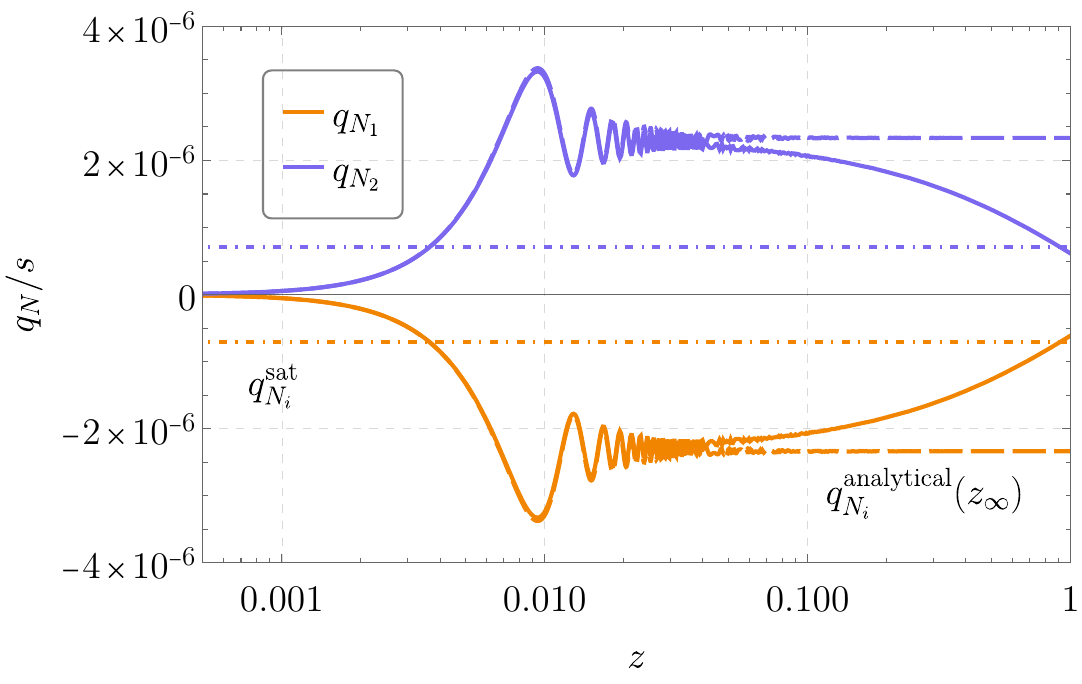}
    \caption{Evolution of the sterile charges with respect to $z$ for the benchmark point. The dashed lines denote the analytical approximations $q_{N_{i}}^{\text{analytical}}(z)$ discussed in the main text, while the dash-dotted lines indicate the final values $q_{N_i}^{\text{sat}}$.}
    \label{fig: sterile charges}
\end{figure}
This analytical approximation is shown as dashed lines in Fig.~\ref{fig: asymmetry vs Yukawa}. To illustrate the generation of the sterile charges, Fig.~\ref{fig: sterile charges} shows the evolution of $q_{N_i}$ with respect to $z$ for the benchmark point discussed in the last section. The solid lines represent the full numerical solutions, while the dashed lines correspond to the analytical solutions given in Eq.~\eqref{eq: analytical solution for the charges}. We observe that the analytical approximation accurately matches the numerical results up to $z \approx 5 \times 10^{-2}$, beyond which the SM back reaction becomes important and the charges decay. The decay is well approximated by the solution in Eq.~\eqref{eq: general matrix solution} where we have used $z_{in} \rightarrow \infty$ such that the initial conditions $\Delta_a^{\text{osc}}$ and $q_{N_i}^{\text{analytical}}$ saturate at a constant value. The solutions to Eq.~\eqref{eq: general matrix solution} evaluated at $z=1$ are shown as dash-dotted lines and labeled as $q_{N_i}^{\text{sat}}$ in Fig.~\ref{fig: sterile charges}.

\begin{figure}
    \centering
    \includegraphics[width=0.7\linewidth]{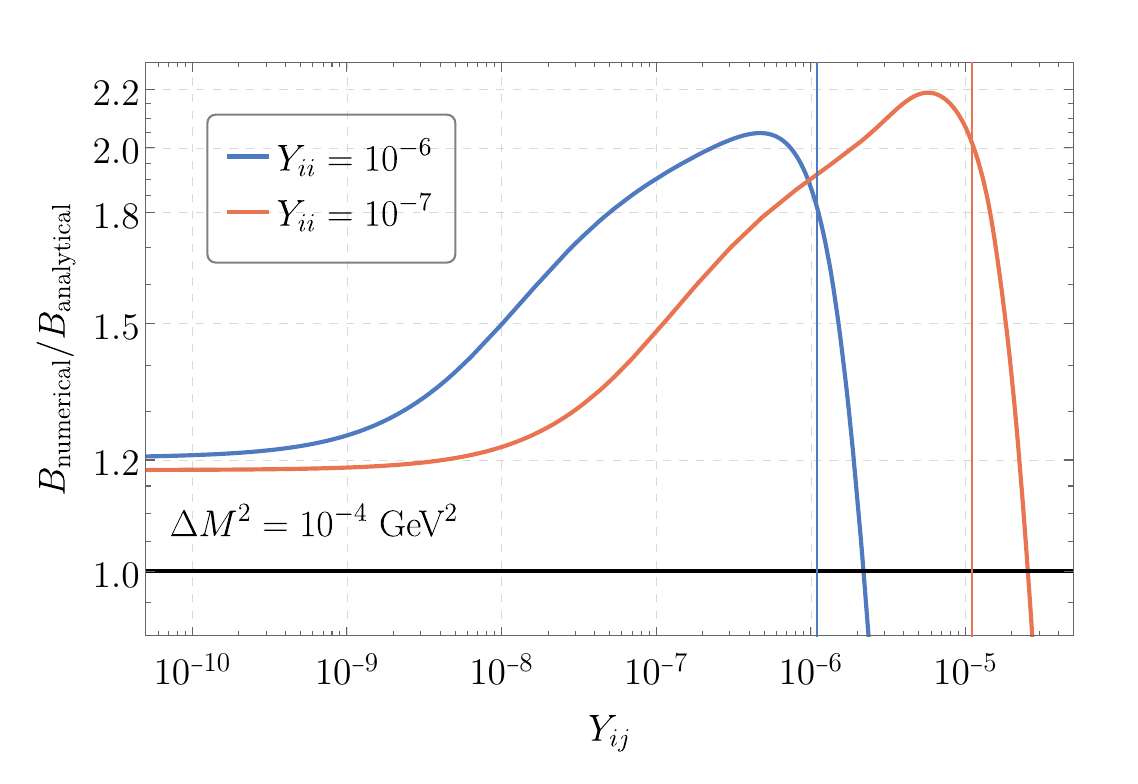}
    \caption{Evolution of the ratio of the full numerical solution and the analytical approximation with respect to the off-diagonal Yukawa elements for different values of the diagonal elements $Y_{ii}=10^{-6}$ (blue) and $Y_{ii} = 10^{-7}$ (orange) for a fixed mass splitting $\Delta M^2 = 10^{-4}~ \text{GeV}^2$. The vertical lines indicate the point at which we expect the analytical approximation to break down.}
    \label{fig: comparison numerical vs analytical}
\end{figure}

As shown in Fig.~\ref{fig: asymmetry vs Yukawa}, however, Equation~\eqref{eq: general matrix solution} does not fully capture the system's behavior across the entire parameter space. As discussed earlier in this section, neglecting the effects of the new physics rate is only valid for Yukawa couplings $Y_{ij} \lesssim 2 \times 10^{-6}$ when $Y_{ii} = 10^{-6}$. Moreover, our approximation only holds for $z_{\text{osc}} < z_{\text{pot}}$ which leads to the condition
\begin{equation}
    Y_{ij} \leq \left(\frac{216 \zeta(3)}{\pi a_R}\right)^{2/3}\frac{(\Delta M^2)^{1/3}}{Y_{ii}} \approx 2.4 \times 10^{-11}\frac{(\Delta M^2)^{1/3}}{Y_{ii}}. 
    \label{eq: break down condition}
\end{equation}
We set the equality and represent this value as vertical lines in Fig.~\ref{fig: asymmetry vs Yukawa}. We expect our approximation to break down for larger values of $Y_{ij}.$ To better visualize the performance of Eq.~\eqref{eq: general matrix solution}, Figure~\ref{fig: comparison numerical vs analytical} finally shows the evolution of the ratio between the full numerical and the analytical solutions as a function of the off-diagonal $Y_{ij}$. We show two fixed values of $Y_{ii} = 10^{-6}$ (blue) and $Y_{ii} = 10^{-7}$ (orange) for $\Delta M^2 = 10^{-4}~\text{GeV}^2$. The vertical solid lines correspond to the upper value of $Y_{ij}$ for which our approximation breaks down as indicated by the condition in Eq.~\eqref{eq: break down condition}. We see that our analytical result differs from the full numerical result by about a factor $\lesssim2$ in the regime of validity and quickly deteriorates outside of it.

\section{Conclusions}
\label{sec: conclusions}

Sterile neutrinos are a very simple and compelling extension of the SM, with the potential to address several major open questions in theoretical physics. They could, for instance, explain the origin of neutrino masses, serve as candidates for dark matter, and account for the baryon asymmetry of the Universe. 
When the masses of the RHNs are below the electroweak scale, leptogenesis can proceed through the ARS mechanism. 
In this scenario, the baryon asymmetry is generated via the CP-violating oscillations of the sterile neutrinos. At the same time, the sterile neutrinos remain out of equilibrium with the SM until after electroweak symmetry breaking and act as reservoir that hides a fraction of the lepton number from the spalerons. 
However, in order to generate the observed BAU, the oscillation timescale, which is set by the mass difference between the sterile neutrinos, cannot be too short compared to the time of sphaleron freeze-out. This necessitates a high degree of degeneracy between the masses, such that GeV sterile neutrinos end up needing to have mass differences of just a few tens of keV.
In this work, we show that adding self-interactions among the RHNs can significantly alleviate this fine-tuned mass condition. To explore this, we employ a scalar-mediated interaction with Yukawa-like couplings to the sterile neutrinos. 
The new interaction has two main effects on the system: 1) it introduces corrections to the neutrino's self-energy and 2) it opens new creation and destruction channels for the sterile neutrinos. This leads to two new time scales that can have an impact on the dynamics. We find that, within the oscillatory regime, the system is dominated by the effects of the thermal potential over a significant portion of the parameter space. In this limit, we can solve the quantum kinetic equations analytically and provide an approximation for the BAU. It agrees with the full numerical solution within a factor of two in its range of validity. Interestingly, we find that the baryon asymmetry can be enhanced by several orders of magnitude due to the new interactions. This relaxes the mass splitting constraint of the vanilla ARS significantly and allows to generate the observed BAU without a strong tuning of the masses. 

Our results show promising new directions to study within the ARS mechanism and more research is needed to explore the possibilities further. In full generality, this is a complicated problem since the different time scales in the problem can arrange in various ways and already within the vanilla ARS mechanism other regimes with alternative solutions are known to exist, see e.g. the discussion of the strongly overdamped regime in \cite{Drewes:2016gmt}. We have not attempted to explore these alternative regions of the parameter space systematically and it remains an interesting question what effects the new self-interactions can have there. In addition, we have applied some simplifications to the system to reduce the number of degrees of freedom. 
For instance, this paper focuses on the case in which the diagonal elements of the Yukawa matrix are equal. Exploring the more general case with $Y_{ii} \neq Y_{jj}$ remains a task for future work. Finally, a comprehensive phenomenological study would be highly desirable to connect the effects of particle dynamics in the early Universe with observables that can be tested in the laboratory today,  e.g. direct and indirect searches for heavy sterile neutrinos (see \cite{Drewes:2015iva} for a summary) or searches for new light scalars at the LHC.

\appendix
\section{The Casas-Ibarra parametrization}
\label{app: Casa-Ibarra parametrization}

For the system considered in this work, the Yukawa matrix $F$ appearing in Eq.~\eqref{eq: lagrangian} can be written using the Casas-Ibarra parametrization \cite{Casas:2001sr} 

\begin{equation}
    F^{\dagger} = \frac{1}{v} U_{\nu} \sqrt{m^{\text{diag}}}  \mathcal{R} \sqrt{M^{\text{diag}}},
    \label{eq: yukawa coupling}
\end{equation}
where $v = 246 ~\text{GeV}$ is the vacuum expectation value of the Higgs and the diagonal matrices $m$ and $M$ contain the masses for the active and the sterile neutrinos, respectively. In the same way $U_{\nu}$ is the PMNS matrix that we parametrize as 
\begin{equation}
   U_{\nu} =  \begin{pmatrix}
         c_{12}c_{13} & c_{13} s_{12} & s_{13} e^{- i \delta} \\ -c_{23} s_{12} - c_{12} e^{i \delta} & c_{12}c_{23} - s_{12} s_{13}s_{23}e^{i \delta} & c_{13} s_{23} \\ -c_{12}c_{23}s_{13}e^{i \delta} + s_{12}s_{23}& -c_{23}s_{12} s_{13} e^{i \delta} - c_{12}s_{23} & c_{13} c_{23}
    \end{pmatrix} \cdot \begin{pmatrix}
        e^{\frac{i}{2}  \alpha_1 } &0&0 \\ 0&1&0  \\ 0&0& e^{\frac{i}{2} \alpha_2}
    \end{pmatrix},
\end{equation}
where for readability we have used the notation $s_{ij} = \sin(\theta_{ij})$ and $c_{ij} = \cos(\theta_{ij})$. In this parametrization, $\theta_{ij}$ are the so-called mixing angles, while $\delta$ is a Dirac phase and $\alpha_{i,j}$ are Majorana phases. In analogy with the PMNS matrix for the active sector, the matrix $\mathcal{R}$ measures the misalignment between the mass and the interaction eigenstates of the sterile neutrinos. For two sterile neutrinos and normal hierarchy, it takes the form 
\begin{equation}
    \mathcal{R} = \begin{pmatrix}
        0 & 0 \\ \cos(\omega) & \sin(\omega) \\ - \sin(\omega) &  \cos(\omega)
    \end{pmatrix},
\end{equation}
with $\omega$ a complex angle so that $\mathcal{R}$ is orthogonal.  Not all input parameters are on the same footing here. On the one hand, the mixing angles for the PMNS matrix are well determined by observations and its CP phases are at least mildly constrained. On the other hand, $\omega$ is currently completely undetermined. It parametrizes our ignorance of the details of the Yukawa couplings of the sterile neutrinos.

\section{Analytical approach to standard ARS}
\label{sec: Garbrecht analytical approach}

In this appendix we review the analytical solution for the ARS scenario in the oscillatory regime. Analytical solutions in this regime have been put forward in, e.g. Refs.~\cite{Asaka:2005pn, Drewes:2012ma, Abada:2015rta, Hernandez:2015wna}. Here, we focus on the solution of Ref.~\cite{Drewes:2016gmt} and we refer the reader to this reference for more details.

Let us start by considering the system at early times, for $z \sim z_{\text{osc}}.$ The condition $z_{\text{osc}} \ll z_{\text{eq}}$ allows us to simplify Eq.~\eqref{eq: integrated Boltzmann eq} in two ways. First, we can neglect the backreaction term from the SM driven by $\tilde{\Gamma}$, since it only becomes important at later times for $z \gg z_{\text{osc}}$. Thus Eq.~\eqref{eq: integrated Boltzmann eq} reduces to \footnote{Given that $H_{N^{\pm}}^{\text{th}}$ is generated by forward scattering of neutrinos through the $F$ couplings, it is diagonal in the flavour basis. This means that, at early times, $H_{N^{\pm}}^{\text{th}}$  commutes with $\delta n_{N^{\pm}}$ and therefore we neglected its contribution in Eq.~\eqref{eq: simplified integrated boltzmann eq}.}
\begin{equation}
    \frac{d}{dz}\delta n_{N^{\pm}} +  \frac{i}{2} z^2\left[ H_{N^{\pm}}^{\text{vac}}, \delta n_{N^{\pm}}\right]= - \frac12 \left\{\Gamma_{N^{\pm}}, \delta n_{N^{\pm}}\right\}.
    \label{eq: simplified integrated boltzmann eq}
\end{equation}
Note that, in the following, we only focus on the equations for $\delta n_{N_{ij}}^{\text{odd/even}}$ and $\delta n^{\text{odd}}_{N_{ii}}$ since the diagonal even components are not relevant for the generation of the BAU. Secondly, we can perform an iterative solution for small $|F^* F^T|$. In this way, the first non-vanishing contributions  can be obtained by replacing $(\delta n_{N^{\pm}})_{ij} \rightarrow - n_{\text{eq}} \delta_{ij}$ on the right-hand side of Eq.~\eqref{eq: integrated Boltzmann eq}. With these simplifications we are left with two equations for the off-diagonal components 
\begin{align}
    \frac{d}{dz} \delta n_{N_{ij}}^{\text{odd}} + i \Omega_{ij} z^2 \delta n_{N_{ij}}^{\text{odd}} & = -i \text{Im}[F^* F^T]_{ij} G , \nonumber \\  \frac{d}{dz} \delta n_{N_{ij}}^{\text{even}} + i \Omega_{ij} z^2 \delta n_{N_{ij}}^{\text{even}} & =  \text{Re}[F^* F^T]_{ij} G,
    \label{eq: Garbrecht eq}
\end{align}
 where 
 \begin{equation}
     \Omega_{ij} = \frac{a_R \pi^2}{T_{\text{ws}}^3 36 \zeta(3)} \left( M_{ii}^2 - M_{jj}^2\right), \quad \text{and} \quad G = \gamma_{AV} \frac{a_R n_{\text{eq}}}{T_{\text{ws}}}.
     \label{eq: definition of omega and G}
 \end{equation}
 The solutions to Eqs.~\eqref{eq: Garbrecht eq} are given by 
 \begin{equation}
     \delta n_{N_{ij}}^{\text{odd}} = -i \text{Im}[F^* F^T] G \mathcal{F}_{ij}, \quad \text{and} \quad \delta n_{N_{ij}}^{\text{even}} = \text{Re}[F^*F^T] G \mathcal{F}_{ij},
     \label{eq: zeroth- order solution}
 \end{equation}
 in terms of the function $\mathcal{F}_{ij}$ which is defined as 
 \begin{equation}
     \mathcal{F}_{ij} = \exp\left(-\frac{1}{3} i \Omega_{ij} z^3\right) \left[\frac{\Gamma\left(\frac13\right) - \Gamma\left(\frac13 , -\frac13 i \Omega_{ij} z^3 \right)}{3^{2/3} \left( -i \Omega_{ij}\right)^{1/3}}\right].
 \end{equation}
Similarly, the evolution of sterile charges, given by the odd diagonal components, is described by 
\begin{align}
    \frac{d}{dz} \delta n_{N_{ii}}^{\text{odd}} &= - \gamma_{AV} \frac{a_R}{T_{\text{ws}}} \Biggl[\text{Re}[F^*F^T]_{ii}\delta n_{N_{ii}}^{\text{odd}} +  \nonumber \\  &  \sum_{\substack{j\\j \neq i}} \left( \text{Re}[F^*F^T]_{ij} \text{Re}[\delta n_{N_{ij}}^{\text{odd}}] + \text{Im}[F^*F^T]_{ij} \text{Im}[\delta n_{N_{ij}}^{\text{even}}]\right)\Biggr]. 
    \label{eq: zeroth order sterile charges}
\end{align}

It follows from Eq.~\eqref{eq: zeroth- order solution} that the source term in Eq.~\eqref{eq: zeroth order sterile charges} vanishes and no sterile charges are generated at this level. Consequently, in the standard ARS mechanism, a net baryon asymmetry is established only at later times, at order $\mathcal{O}(F^6)$, due to the washout effects from the SM. Nonetheless, flavour asymmetries of order $\mathcal{O}(F^4)$ are produced already at $z_{\text{osc}}$ through the source term in Eq.~\eqref{eq: source term} 

\begin{equation}
    \Delta^{\text{osc}}_a (z) = - \int_0^z \frac{dz'}{T_{\text{ws}}} S_a,  
    \label{eq: Garbrect initial value flavour charges}
\end{equation}
which, at this stage, satisfy $\sum_{\substack{a = e, \mu, \tau}} \Delta_a = 0 $. 

Once $z \sim z_{\text{eq}}\gg z_{\text{osc}}$ the oscillations of the off-diagonal correlations \eqref{eq: zeroth- order solution} are too fast and they average to zero. So the evolution of the flavour asymmetries and the sterile charges at this later stage can be described by Eqs.~\eqref{eq: late time evolution}. The solution is then given by Eq.~\eqref{eq: general matrix solution}, where now we take as initial conditions $q_{N_i}(z_{in}) = 0$ and 
\begin{equation}
    \Delta_a^{\text{osc}} (z_{in}) = -\frac{i z^2}{2 T_{\text{ws}}}a_R~ \gamma_{AV}~ G \left.\left(\sum_{\substack{i,j,c \\ i\neq j}} F^{\dagger}_{ai} F_{ic} F^{\dagger}_{cj} F_{ja} \text{Im}\left[{_2}F_2 \left[\left\{ \frac23 , 1 \right\} ; \left\{ \frac43, \frac53\right\}; -\frac13 i z^3 \Omega_{ij}\right]\right]\right)\right|_{z\rightarrow \infty},
\end{equation}
which is the analytical solution to Eq.\eqref{eq: Garbrect initial value flavour charges}.

\section{Derivation of the quantum kinetic equation}
\label{app: derivation of the QKE}

In this appendix, we derive the terms associated with sterile neutrino self-interactions in Eq.~\eqref{eq: Full QKE new physics}:
\begin{equation}
\frac{d \rho}{dt} = -i \left[H^{\text{th}}_{\phi},\rho\right] - \rho ~ \Gamma^d_{\phi} ~\rho + \Gamma_{\phi}^p.
\label{eq: QKE new physics}
\end{equation}
Our derivation follows a diagrammatic approach, as outlined in e.g.~\cite{Brambilla:2017zei}. The appropriate theoretical setting for studying interacting particles in a thermal environment is thermal field theory, with the real-time formalism (also known as closed-time path formalism) offering a natural framework for addressing time-dependent quantities. In this formalism, the finite-temperature fermion propagator acquires a matrix structure which, in momentum space, reads  
\begin{equation}
   S^T (k) = \begin{pmatrix}
        S^{++}(k) & S^{+-}(k) \\ S^{-+}(k) & S^{--}(k)
    \end{pmatrix},
    \label{eq: matrix form of the propagator}
\end{equation}
with 
\begin{equation}
    iS^{++}(k) = (iS^{--}(k))^* = (\slashed{k} +m )\left[ \frac{i}{k^2 -m^2 +i \epsilon }  - 2\pi \delta(k^2-m^2) f_f(|k_0|)\right]
\end{equation}
the physical thermal propagator consisting of the vacuum Feynman propagator plus a thermal correction. The off-diagonal components read 
\begin{align}
    S^{+-}(k) &= -2\pi i(\slashed{k} +m ) \delta(k^2-m^2) f_f(|k_0|),\\ 
    S^{-+}(k) &= -2\pi i(\slashed{k} +m )\delta(k^2 -m^2) [1- f_f(|k_0|)],
\end{align}
where $f_f$ is the corresponding fermionic distribution function (see \cite{Bellac:2011kqa} for a careful derivation). In a similar way, the scalar propagator components are given by 
\begin{align}
    iD^{++}(k) &= (iD^{--}(p))^* = \frac{i}{k^2 - m^2 +i\epsilon} + 2\pi \delta(k^2-m^2)f_b(|k_0|), \\
    D^{+-} (k) &= 2\pi i \delta(k^2-m^2) f_b(|k_0|), \\ 
    D^{-+} (k) &= 2\pi i \delta(k^2-m^2)[1 + f_b(|k_0|)],  
\end{align}
with $f_b$ a bosonic distribution function. Analogously to the fermionic case, the ($++$) component comprises the physical propagator for the scalar. In our case at hand, in a system with two neutrinos, the components of the fermion propagator carry flavor indices and, therefore, become matrix-valued in flavor space. In particular, we are interested in the evolution of $S^{+-}_{ij}(k)$ which is related to the density matrix as 
\begin{equation}
     S_{ij}^{+-} (k) = -2\pi i \delta(k^2) \slashed{k} ~\rho_{ij}(k), 
\end{equation}
where we have assumed massless neutrinos. 
Here $i,j$ are flavor indices and the sum is over all helicity states (positive and negative). The ($++$) and ($--$) components of the propagator remain diagonal in flavor space. In order to find an equation for the density matrix it is convenient to work in coordinate space where \begin{equation}
S_{ij}^{+-}(t,\vec 0; t, \vec0) = i \langle \bar{\psi}_j(t,\vec0) \psi_i(t,\vec0)\rangle = i \rho_{ij}(t;t)\end{equation}
with $\psi_j(t,\vec0)$ Majorana neutrino fields. Diagrammatically we can then write
$$\vcenter{\hbox{
{\begin{tikzpicture}
\begin{feynman}
\vertex (a1) {};
\vertex[above =0.3cm of a1] (a2){\(i\)};
\vertex[below right =0.4cm of a1] (a3){\(\pmb{+}\)};
\vertex[right=3.0cm of a1] (a4) ;
\vertex[ right= 3.2cm of a2] (a5){\(j\)};
\vertex[right = 2.6 cm of a3] (a6){\(\pmb{-}\)};
\diagram* {
{[edges=fermion]
(a1)  --  (a4),
},
};
\end{feynman}
\end{tikzpicture}}
}}  = iS^{+-}_{ij} = e^{-iH_0 (t-t_0)}  \rho_{ij}(t_0;t_0) e^{iH_0(t-t_0)},$$
where the `\(\pmb{+}\)' and `\(\pmb{-}\)' symbols below the propagator indicate insertions of fields from the upper or lower branches of the closed-time path \cite{Bellac:2011kqa}, i.e. the corresponding components in Eq.\eqref{eq: matrix form of the propagator}. Also, note that we work in the interaction picture so the time evolution of the density matrix is governed by the free Hamiltonian $H_0$. We now proceed to include corrections to this propagator up to second order in $\rho_{ij}.$    Schematically this is  
\begin{align}
    & \hspace{0.7cm} \vcenter{\hbox{
{\begin{tikzpicture}
\begin{feynman}
\vertex (a1) {};
\vertex[above =0.3cm of a1] (a2){\(i\)};
\vertex[below right =0.4cm of a1] (a3){\(\pmb{+}\)};
\vertex[right=3.0cm of a1] (a4) ;
\vertex[ right= 3.2cm of a2] (a5){\(j\)};
\vertex[right = 2.6 cm of a3] (a6){\(\pmb{-}\)};
\diagram* {
{
(a1)  --  (a4),
},
};
\end{feynman}
\end{tikzpicture}}}}
 = \nonumber \\ &  
\quad \quad  \vcenter{\hbox{
{\begin{tikzpicture}
\begin{feynman}
\vertex (a1) {};
\vertex[above =0.3cm of a1] (a2){\(i\)};
\vertex[below right =0.4cm of a1] (a3){\(\pmb{+}\)};
\vertex[right=6.6cm of a1] (a4) ;
\vertex[ right= 6.6cm of a2] (a5){\(j\)};
\vertex[right = 6.2 cm of a3] (a6){\(\pmb{-}\)};
\vertex[right = 3.35 cm of a1](blob);
\diagram* {
{[]
(a1)  --  (a4),
},
};
\draw[fill=black] (blob) circle (3pt);
\end{feynman}
\end{tikzpicture}} 
}}  
+ ~\raisebox{-0.5cm}{{\begin{tikzpicture}
\begin{feynman}
\vertex (a1) ;
\vertex[right=2cm of a1] (a2);
\vertex[above =0.1cm of a1] (a21){\(i\)};
\vertex[below right =0.1cm of a1] (a31){\(\pmb{+}\)};
\vertex[right= 1.6cm of a31]{\(\pmb{-}\)};
\vertex[right= 4.2cm of a31]{\(\pmb{-}\)};
\vertex[right=2.5cm of a2] (a3);
\vertex[right=2.0cm of a3] (a4);
\vertex[ right= 6.5cm of a21] (a5){\(j\)};
\vertex[right = 5.9 cm of a31] (a6){\(\pmb{-}\)};
\vertex[right = 1 cm of a1](blob);
\vertex[right = 5.5 of a1](number);
\vertex[above = 1cm of number](number1);
\diagram* {
{
(a1) -- [] (a2) -- [] (a3) --  (a4),
},
(a2) -- [scalar,out=90, in=90, looseness=1.5, edge label=\(\phi\)] (a3),
};
\draw[fill=black] (blob) circle (3pt);
\node[draw, circle, inner sep=2pt] at (number1) {\textsc{I}};
\end{feynman}
\end{tikzpicture}}
}\nonumber \\  &\quad + ~\raisebox{-0.5cm}{{\begin{tikzpicture}
\begin{feynman}
\vertex (a1) ;
\vertex[right=2cm of a1] (a2);
\vertex[above =0.1cm of a1] (a21){\(i\)};
\vertex[below right =0.1cm of a1] (a31){\(\pmb{+}\)};
\vertex[right= 1.6cm of a31]{\(\pmb{+}\)};
\vertex[right= 4.2cm of a31]{\(\pmb{+}\)};
\vertex[right=2.5cm of a2] (a3);
\vertex[right=2.0cm of a3] (a4);
\vertex[ right= 6.5cm of a21] (a5){\(j\)};
\vertex[right = 5.9 cm of a31] (a6){\(\pmb{-}\)};
\vertex[right = 5.5 cm of a1](blob);
\vertex[above = 1cm of blob](number1);
\diagram* {
{
(a1) -- [] (a2) -- [] (a3) --  (a4),
},
(a2) -- [scalar,out=90, in=90, looseness=1.5, edge label=\(\phi\)] (a3),
};
\draw[fill=black] (blob) circle (3pt);
\node[draw, circle, inner sep=2pt] at (number1) {\textsc{II}};
\end{feynman}
\end{tikzpicture}}
}  + ~\raisebox{-0.5cm}{{\begin{tikzpicture}
\begin{feynman}
\vertex (a1) ;
\vertex[right=2cm of a1] (a2);
\vertex[above =0.1cm of a1] (a21){\(i\)};
\vertex[below right =0.1cm of a1] (a31){\(\pmb{+}\)};
\vertex[right= 1.6cm of a31]{\(\pmb{+}\)};
\vertex[right= 4.2cm of a31]{\(\pmb{-}\)};
\vertex[right=2.5cm of a2] (a3);
\vertex[right=2.0cm of a3] (a4);
\vertex[ right= 6.5cm of a21] (a5){\(j\)};
\vertex[right = 5.9 cm of a31] (a6){\(\pmb{-}\)};
\vertex[right = 3.25 cm of a1](blob);
\vertex[right = 5.5 of a1](number);
\vertex[above = 1cm of number](number1);
\diagram* {
{
(a1) -- [] (a2) -- [] (a3) --  (a4),
},
(a2) -- [scalar,out=90, in=90, looseness=1.5, edge label=\(\phi\)] (a3),
};
\draw[fill=black] (blob) circle (3pt);
\node[draw, circle, inner sep=2pt] at (number1) {\textsc{III}};
\end{feynman}
\end{tikzpicture}}
} \nonumber \\& \quad + ~\raisebox{-0.5cm}{{\begin{tikzpicture}
\begin{feynman}
\vertex (a1) ;
\vertex[right=2cm of a1] (a2);
\vertex[left =0.1cm of a1] (a21){\(i\)};
\vertex[above =0.2cm of a1](time){\(t\)};
\vertex[right =1.7cm of time]{\(t_2\)};
\vertex[right =4.8cm of time]{\(t_1\)};
\vertex[right =6.5cm of time]{\(t\)};
\vertex[below right =0.1cm of a1] (a31){\(\pmb{+}\)};
\vertex[right= 1.6cm of a31]{\(\pmb{-}\)};
\vertex[right= 4.2cm of a31]{\(\pmb{+}\)};
\vertex[right=2.5cm of a2] (a3);
\vertex[right=2.0cm of a3] (a4);
\vertex[ right= 7.1cm of a21] (a5){\(j\)};
\vertex[right = 5.9 cm of a31] (a6){\(\pmb{-}\)};
\vertex[right = 1 cm of a1](blob);
\vertex[right = 5.5 cm of a1](blob1);
\vertex[above = 1cm of blob1](number1);
\diagram* {
{
(a1) -- [] (a2) -- [] (a3) --  (a4),
},
(a2) -- [scalar,out=90, in=90, looseness=1.5, edge label=\(\phi\)] (a3),
};
\draw[fill=black] (blob) circle (3pt);
\draw[fill=black] (blob1) circle (3pt);
\node[draw, circle, inner sep=2pt] at (number1) {\textsc{IV}};
\end{feynman}
\end{tikzpicture}}
} + ~\mathcal{O}(\rho^3),
\label{eq: schematic QKE}
\end{align}
where the blobs indicate the legs associated with the density matrix, which correspond to the ($+-$) component of the propagator. In the last loop diagram we also include the time labels that we use below.   
\begin{align}
    \raisebox{-0.5cm}{{\begin{tikzpicture}
\begin{feynman}
\vertex (a1) ;
\vertex[right=1cm of a1] (a2);
\vertex[above =0.1cm of a1] (a21){\(i\)};
\vertex[right = 0.8 cm of a21]{\(k\)};
\vertex[right = 1.4 cm of a21]{\(m\)};
\vertex[right = 2.7 cm of a21]{\(l\)};
\vertex[below =0.1cm of a1] (a31){\(\pmb{+}\)};
\vertex[right= 1 cm of a31]{\(\pmb{-}\)};
\vertex[right= 3cm of a31]{\(\pmb{-}\)};
\vertex[right=2cm of a2] (a3);
\vertex[right=1.0cm of a3] (a4);
\vertex[ right= 4cm of a21] (a5){\(j\)};
\vertex[right = 4 cm of a31] (a6){\(\pmb{-}\)};
\vertex[right = 0.5 cm of a1](blob);
\vertex[right = 5.5 of a1](number);
\vertex[above = 1cm of blob](number1);
\diagram* {
{
(a1) -- [] (a2) -- [] (a3) --  (a4),
},
(a2) -- [scalar,out=90, in=90, looseness=1.5, edge label=\(\phi\)] (a3),
};
\draw[fill=black] (blob) circle (3pt);
\node[draw, circle, inner sep=2pt] at (number1) {\textsc{I}};
\end{feynman}
\end{tikzpicture}}
} &= \int_{t_0}^t dt_1~ e^{-iH_0 (t-t_1)} \int_{t_0}^{t_1} dt_2 ~e^{-iH_0 (t_1-t_2)}  (-i Y_{jl})~iS^{--}_{lm}(t_1-t_2) \delta_{lm}  \nonumber \\  & \quad iD^{--}(t_1-t_2)(-i Y_{mk}) e^{-iH_0(t_2-t_0)}  \rho_{ik}(t_0;t_0) e^{iH_0(t-t_0)}.  
\end{align}
Here we use Einstein's summation convention for repeated indices. Defining
\begin{equation}
    \Sigma_{jk}(t_1) = \int_{t_0}^{t_1}dt_2~ e^{-iH_0(t_1-t_2)} Y_{jm} [iS^{++}_{mm}(t_1-t_2)] [iD^{++}(t_1-t_2)]~ Y_{mk}~ e^{iH_0 (t_1-t_2)},
\end{equation}
we can rewrite the first diagram as 
\begin{align}
    \raisebox{-.1cm}{{\begin{tikzpicture}
\begin{feynman}
\vertex(number1);
\node[draw, circle, inner sep=2pt] at (number) {\textsc{I}};
\end{feynman}
\end{tikzpicture}}
} &= - \int_{t_0}^{t} dt_1 ~e^{-iH_0(t-t_1)} \Sigma_{kj}^{\dagger}(t_1) e^{-iH_0(t_1-t_0)} \rho_{ik}(t_0;t_0) e^{i H_0(t-t_0)} =   -\int_{t_0}^{t} dt_1 ~e^{H_0(t-t_1)} \Sigma_{kj}^{\dagger}(t_1)  \rho_{ik}(t_1;t),
\end{align} 
Similarly, we find 
\begin{align}
    \raisebox{-.1cm}{{\begin{tikzpicture}
\begin{feynman}
\vertex(number1);
\node[draw, circle, inner sep=2pt] at (number) {\textsc{II}};
\end{feynman}
\end{tikzpicture}}
} &=  - \int_{t_0}^{t} dt_1 ~\rho_{kj}(t;t_1) \Sigma_{ik}(t_1) e^{-iH_0(t_1-t)}, \nonumber  \\ \raisebox{-.1cm}{{\begin{tikzpicture}
\begin{feynman}
\vertex(number1);
\node[draw, circle, inner sep=2pt] at (number) {\textsc{III}};
\end{feynman}
\end{tikzpicture}}
} &=  \int_{t_0}^{t} dt_1~ e^{-iH_0 (t-t_1)}~ \Xi (\rho_0,t_1) ~e^{-iH_0(t_1-t)},\nonumber \\ \raisebox{-.1cm}{{\begin{tikzpicture}
\begin{feynman}
\vertex(number1);
\node[draw, circle, inner sep=2pt] at (number) {\textsc{IV}};
\end{feynman}
\end{tikzpicture}}
} &= - \int_{t_0}^t dt_1~ \rho_{jn}(t;t_1)~ \Xi^d_{nk}(t_1)~ \rho_{ki}(t_1;t),
\end{align}
with 
\begin{align}
    \Xi(\rho_0,t_1) &= \int_{t_0}^{t_1} dt_2~ Y_{nl} ~  \rho_{\phi} e^{-iH_0(t_1-t_0)} \rho_{lm}(t_0;t_0) e^{iH_0(t_2-t_0)} Y_{mk} ~e^{iH_0(t_1-t_2)}, \nonumber \\ 
    \Xi^d_{nk}(t_1) &= \int_{t_0}^{t_1} dt_2~ e^{-iH_0(t_1-t_2)}Y_{nm} Y_{mk}~ e^{iH_0(t_1-t_2)}.
\end{align}
Here $\rho_\phi$ is the distribution function of the scalar that we assume is in equilibrium with the SM plasma. Taking the derivative on both sides of Eq.~\eqref{eq: schematic QKE} gives 
\begin{align}
    \frac{d\rho(t;t)}{dt} = -i H_0 \rho(t;t) + i H_0 \rho(t;t) - \rho(t;t)\Sigma(t)^\dagger - \Sigma(t) \rho(t;t) + \Xi(\rho_0, t) - \rho(t;t)~ \Xi^d(t) \rho(t;t).  
\end{align}
The interpretation of the functions appearing above is now manifest. In particular, we recognize the sterile neutrinos' self-energy as $-i\Sigma$ which provides in-medium masses and widths. Using that   
\begin{align}
   \begin{cases}
       -i \Sigma(t) + i \Sigma^\dagger(t) = 2 \text{Re}(-i\Sigma) \equiv 2 V_\phi,\\ 
       \Sigma(t)+ \Sigma^\dagger(t) = -2 \text{Im}(-i\Sigma) \equiv \Gamma_s,
   \end{cases} 
\end{align}
where we identify the real and imaginary parts of the self-energy with the effective potential $V_\phi$ and the thermal width $\Gamma_s$ of the sterile neutrinos respectively, the QKE reduces to a more familiar form  
\begin{equation}
    \frac{d\rho}{dt} = -i \left[H_0,\rho\right] - i \left[V_\phi,\rho\right] - \frac12 \left\{\Gamma_s,\rho\right\}+ \Xi - \rho~ \Xi^d \rho. 
\end{equation}
In our case at hand, the width of the sterile neutrinos vanishes, since the decay into a heavier scalar is kinematically forbidden. Moreover, identifying the functions $\Xi$ and $\Xi^d$ as the production and destruction rates respectively, we finally obtain the desired result in Eq.\eqref{eq: QKE new physics}.      

\acknowledgments

MD and SV acknowledge support by the DFG via the the individual research grant Nr. 496940663. We thank S. Biondini for helpful discussions on thermal field theory.









\bibliographystyle{JHEP.bst}
\bibliography{freeze-in_leptogenesis.bib}
\end{document}